# Energy of adsorption at semiconducting surfaces with Fermi level differently pinned - ab initio study

Paweł Kempisty[1], Paweł Strak[1], Konrad Sakowski[1], Stanisław Krukowski[1,2]

[1]Institute of High Pressure Physics, Polish Academy of Sciences, Sokołowska 29/37, 01-142 Warsaw, Poland

[2]Interdisciplinary Centre for Modelling, University of Warsaw, Pawińskiego 5a, 02-106 Warsaw, Poland

It is shown that adsorption energy at semiconductor surfaces critically depends on the charge transfer to or from the adsorbed species. For the processes without the charge transfer, such as molecular adsorption of closed shell systems, the adsorption energy is determined by the bonding only. In the case involving charge transfer, such as open shell systems like metal atoms or the dissociating molecules, the energy attains different value for the Fermi level differently pinned. The DFT simulation of adsorption of ammonia at hydrogen covered GaN(0001) confirms these predictions: the molecular adsorption is independent on the coverage while the dissociative process energy varies by several electronvolts.

# I. Introduction

Adsorption of various species at the solid surfaces could be divided based on the properties of the adsorbed species and also on the type of the material. Two prominent categories of substances are frequently quoted, electric conductors and insulators. Yet within the electrically conductive systems, the metals and semiconductors differ considerably, principally by different bonding type, isotropically cohesive metals and directionally bonded semiconductors. The metal surfaces interact with the active species isotropically and are relatively weakly affected by the adsorption processes.[1,2] Drastically different are semiconductor surfaces where interaction are directional frequently involving reconstructions, the effects that could be drastically changed by the presence of adsorbed species.[3,4]

In addition to the above mentioned chemical and structural properties of their surfaces, the electric properties of the two types of the solids show considerable differences. The metals are immune to the influence by doping as their Fermi level is fixed. Due to drastically high density of the electrons the screening lengths in metals is mere fraction of Angstroms. Therefore the surface properties of metals are successfully simulated by *ab intio* calculations, using either slab or cluster models.[5]

The electric properties of the semiconductor bulk and surfaces are susceptible to manipulation by doping. The Fermi level may be shifted across the whole bandgap, to achieve, in some cases, degeneracy of the electron/hole gas in the conduction/valence bands. Thus, the Fermi energy may be easily shifted by several electronvolts in the case of wide bandgap compounds. The semiconductor surfaces may be electrically charged and are relatively weakly screened that can shift the energy of quantum states in the vicinity of surfaces. Typically, these charges are screened by the mobile or localized charges, over the distances from few Angstroms to microns. The Fermi level maybe free, or pinned by the surface states which causes charge transfer between the surface and the bulk entailing appropriate band bending to adjust to the Fermi energy in the bulk.[3,4]

Such complicated physics of semiconductor surfaces requires simulation procedures more sophisticated than these used for metals. Until very recently the state of art was fairly disappointing. Traditionally, the slab models were employed with the opposite side surface terminated, creating broken bonds saturated by integer or fractional charge pseudohydrogen atoms, satisfying the electron counting (EC) rule.[6] The device, essentially based on simple tight binding arguments was considered to be satisfactory that was supported by the claims that the system properties were size independent.[7-9]

Recently a more developed procedure was devised, based on the observation that manipulation of the surface termination atoms by change of their location or of their charge, the average electric field within the slab may be changed.[10-13] The model was successfully applied for studies of polar surfaces of GaN[10-14] and SiC.[15] The change of the energies of the quantum states by the field at the surface was denoted as the Surface States Stark Effect (SSSE).[11-14] In addition to the surface termination manipulation, an alternative approach was proposed in the recent edition of SIESTA shareware. The second approach, in principle, could be also applied to surface charged states.[15]

In the present study, we analyze the influence of the of Fermi level that may be differently pinned due to different chemical conditions at he surface on the value of adsorption energy. The role of charge transfer will be elucidated using an example of ammonia adsorption at GaN(0001) surface. The molecular adsorption keeps all molecular states of ammonia occupied, thus no charge transfer is possible. The dissociative adsorption leads to the attachment of the separate $NH_2$ radicals and H adatoms with new states emerging at the surface so that charge transfer is possible. The comparison of these two processes allows to pin the basic influence of the charge transfer on the adsorption energy in these processes. The conclusions describe general features of the adsorption processes at semiconductor surfaces, qualitatively different from these at the surfaces of metals.

## II. The simulation procedure

In part of the calculations reported below, a freely accessible DFT code SIESTA, combining norm conserving pseudopotentials with the local basis functions, was employed.[16-19] The functional basis in SIESTA consists of numeric atomic orbitals, of the finite size support, determined by the user. The pseudopotentials for Ga, H and N atoms were generated, using ATOM program for all-electron calculations. SIESTA employs the norm-conserving Troullier-Martins pseudopotential, in the Kleinmann-Bylander factorized form.[20] Gallium 3d electrons were included in the valence electron set explicitly to account specific bonding features of GaN. The following atomic basis sets were used in GGA calculations: Ga (bulk) - 4s: DZ (double zeta), 4p: DZ, 3d: SZ (single zeta), 4d: SZ; Ga (surface)- 4s: TZ (triple zeta), 4p: TZ, 3d: SZ, 4d: SZ; N (bulk) - 2s: TZ, 2p: DZ;  N (surface)- 2s: TZ, 2p: TZ, 3d: SZ; H - 1s: QZ (quadruple zeta), 2p: SZ and H (termination atoms)  1s: DZ, 2p: DZ, 3d: SZ. The following values for the lattice constants of bulk GaN were obtained in GGA-WC calculations (as exchange-correlation functional Wu-Cohen (WC) modification of Perdew, Burke and Ernzerhof (PBE) functional[21,22]: a = b = 3.2021 Å , c = 5.2124 Å. These values are in a good agreement with the experimental data for GaN: a = 3.189 Å and c = 5.185 Å [23]. All the presented dispersion relations are plotted as obtained from DFT calculations, burdened by a standard DFT error in the recovery of GaN bandgap. In the present parameterization, the effective bandgap for bulk GaN was 1.867 eV. Hence, in order to obtain a quantitative agreement with the experimentally measured values, all the calculated DFT energies that were obtained, should be rescaled by an approximate factor $\alpha = E_{g\text{-exp}}/E_{g\text{-DFT}} = 3.4\text{eV}/2.13\text{eV} \approx 5/3 \approx 1.6$. Integrals in k-space were performed using a 3x3x1 Monkhorst-Pack grid for the slab with a lateral size 2x2 unit cell and only Γ-point for 4x4 slabs.[24] As a convergence criterion, terminating a SCF loop, the maximum difference between the output and the input of each element of the density matrix was employed being equal or smaller than $10^{-4}$. Relaxation of the atomic position is terminated when the forces acting on the atoms become smaller than 0.02 eV/Å.

Born-Oppenheimer approximation was used for determination of the energy barriers in which an effective procedure, based on nudged elastic band (NEB) method was applied.[25-27] The NEB method finds minimum energy pathways (MEP) between two stable points, which has to be predetermined first. The MEP is characterized by at least one first order saddle point, finding the energy barrier corresponding to activated energy complex approximation in chemical reaction kinetics. In the present formulation NEB module was linked to SIESTA package paving the way to fast determination of the energy and conformation of the species along the optimized pathways.

## III. Adsorption energy – a role of electron transfer

As discussed previously,[28] in the electrically neutral process, the energy shift due to band bending at the surface is of the same magnitude for negative electron and positive nucleus, thus the net effect for electrically neutral process is zero, i.e. the adsorption energy is identical for p- and n-type material, pinned or not. It was argued also that for the processes involving charge transfer the adsorption energy dependence on the doping of the semiconductor is drastically different for the surface having Fermi level pinned and nonpinned by the surface states in which the difference originates from the energetic effect of the electron transfer between the surface state and the bulk.

For Fermi level pinned by the surface states, the adsorption energy was found to be independent on the doping in the bulk. Since all surface state energies are shifted by the surface band bending equally, therefore the electron transferred from the bulk states at the Fermi level to the emerging surface state gains the energy equal to the energy gain in transfer from the pinning surface state. Thus the energy difference is the same for n- and p-type material as both surface states energies are equally shifted by surface band bending. Therefore the adsorption energy is independent on the doping in the bulk. The simulations of the adsorption of hydrogen confirmed that the adsorption energy is virtually independent on the Fermi level in the sample.[28]

A different scenario is predicted for the case of semiconductor surface with Fermi level not pinned at the surface. The electron transfer between the bulk and the emerging surface state involves the energetic effect that is different for n- and p-type. As in this case the energy of surface state is not correlated with the Fermi energy, the electron transfer energetic effect is different, dependent on the doping in the bulk. The simulations of the adsorption of hydrogen at partially covered GaN(0001) surface proved that the adsorption energy difference between n- and p-type material may be of order of several electronvolts.[28]

In addition to the above considerations, the it is worth to determine whether the adsorption energy is independent of the type of the surface state that pins the Fermi level. As shown in Fig. 1, this issue involves the comparison of the two different chemical coverages of the surface in which the Fermi level is pinned by the two different surface states, i.e. having different energy with respect to valence (valence band maximum - VBM) and conduction bands (conduction band minimum - CBM) at the surface, i.e.

$$E_{1a} = E_{1sa} - E_{VBM} \neq E_{1d} = E_{1sd} - E_{VBM} \qquad (1)$$

In order to demonstrate the possible difference, the most drastic example is presented in Fig. 1 of the first surface state pinning the Fermi level is the acceptor (Fig 1 a) and the second the donor (Fig 1 b) type. In contrary to different Fermi level pinning state, the emerging surface state is identical, thus the energy difference of the state and CBM is identical (denoted as $E_3$ in Fig 1). Still the CBM and surface adsorbed state are affected by the surface band bending differently for the surface donor and for the acceptor (denoted as $e_oV_{sa}$ and $e_oV_{sb}$ respectively). Since, the adsorbing species are electrically neutral, the influence of the potential difference affecting the electrons and nuclei independently, cancels out. Thus the difference of the energy of the electron states far and close at the surface, i.e. $E_{2a} \neq E_{2d}$ has no influence on the adsorption energy.

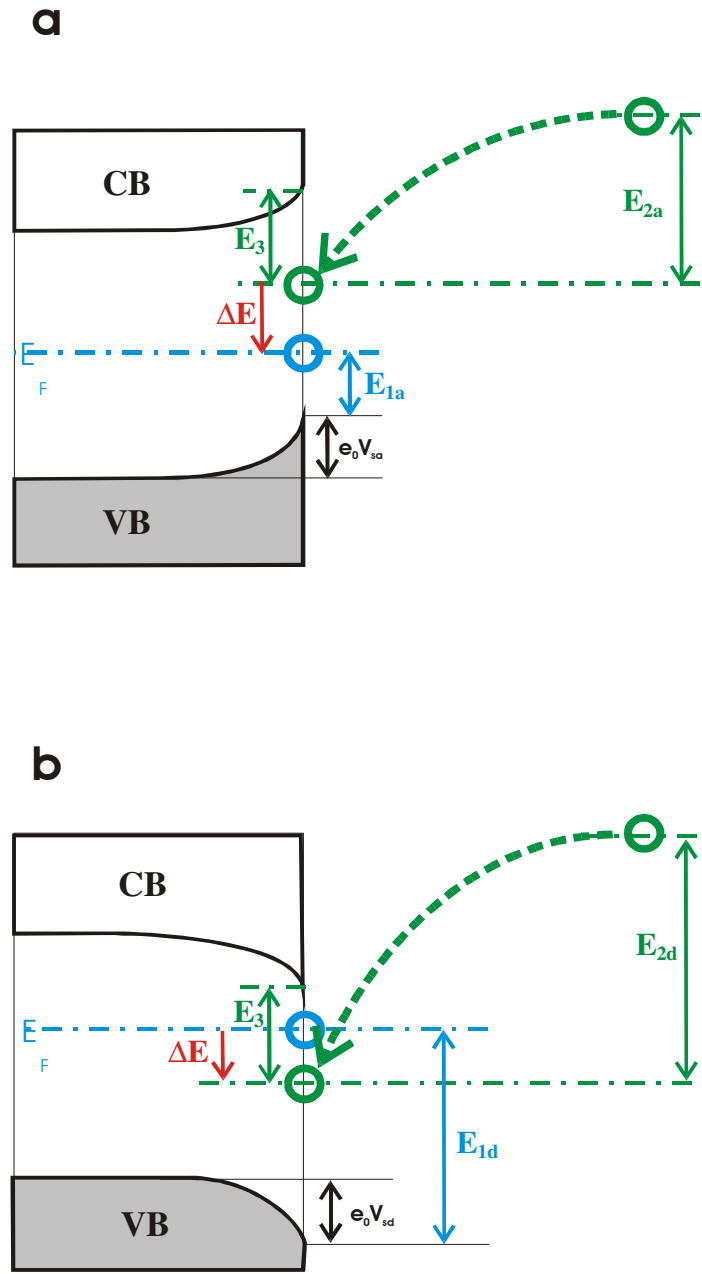

Fig. 1. Energy quantum states, shown in the absolute scale, varying during the adsorption process at the surfaces with Fermi level pinned by the two different surface states: a - surface acceptor, b - surface donor, of their energies $E_{1a}$ and $E_{1d}$ above VBM at the surface, respectively. The change of the energy of electron quantum state of the adsorbing

species $E_{2a}$ and $E_{2d}$, are denoted by green dashed line. Additional contribution to the adsorption energy stems from the charge transfer between the surface states and the bulk states having their energy close to Fermi level, marked by red ($\Delta E$). The energy difference between the pinning surface state and the VBM, equal to $E_{1a}$ and $E_{1d}$, is denoted by blue color.

In the majority of surface adsorption processes the surface state created by the species adsorbed is different from the state pinning Fermi level at the surface. Thus the additional fraction of the adsorption energy stems from the charge transfer to (Fig 1a) or from (Fig 1b) the adsorbed state from/to the full/empty states in the bulk, i.e. their energy close to the Fermi energy. In Fig.1 the difference is denoted as $\Delta E$ (red). In fact the contribution of such transfer to the adsorption energy is different from the difference of single electron energies due to many body effects. It is crucial that such addition may be nonzero only in the case when the quantum states of the adsorbed species allows for such charge transfer. Most likely this phenomenon occurs in the case of open shell systems while closed shell systems are immune to this effect.

The different pinning of Fermi level may be realized most easily by the two different coverages of the same surface. In this case the quantum contribution is identical, thus the dominating term originates from the charge transfer. A notable example is the GaN(0001) surface under small and high hydrogen coverage.[29,30] For small hydrogen coverage, the Fermi level is pinned by Ga broken bond state, located at 0.45 eV below CBM. Since the Ga-H state is degenerate with the valence band, for high hydrogen coverage, the Fermi level is pinned at VBM. Thus the energy difference is about 3 eV, which in DFT implementation it is reduced to about 1.6 eV.

Thus as a working example of the role of pinning surface state, the adsorption of ammonia on H-covered GaN(0001) surface is investigated below, elucidating various contributions to the adsorption energy. The ammonia molecule may be adsorbed in molecular form[31] or dissociated upon adsorption into separate the $NH_2$ radical and the H adatom. The first case plays the role of the closed and the second the open shell system, respectively.[32-34] The species adsorbed in these processes have different bonding that entails zero or finite charge transfer to the surface states. The comparison of the different H-coverage allows to study influence of different pinning states and to verify the description presented above.

## IV. DFT example study - ammonia at GaN(0001) surface

An ammonia molecule is bonded by 8 electrons: 5 originating from nitrogen and 3 from hydrogen atoms. These electrons are distributed on molecular states that are presented in Fig. 2a. In order of increasing energy, these states are: σ - the lowest energy state created from s atomic states, the two σ states created from $p_x$ and $p_y$ atomic states and π state created from $p_z$ atomic states. As discussed in the Introduction, adsorption of ammonia may lead to two different atomic configurations: first in which the entire molecule was attached on the surface, and the second - the dissociation to $NH_2$ radical and H atom, attached at the surface in the bridge and on-top positions, respectively. In both cases the adsorption process leads to drastic transformation of the part of ammonia molecular states. These two positions and these new states, preserved and transformed are presented in Fig. 3.

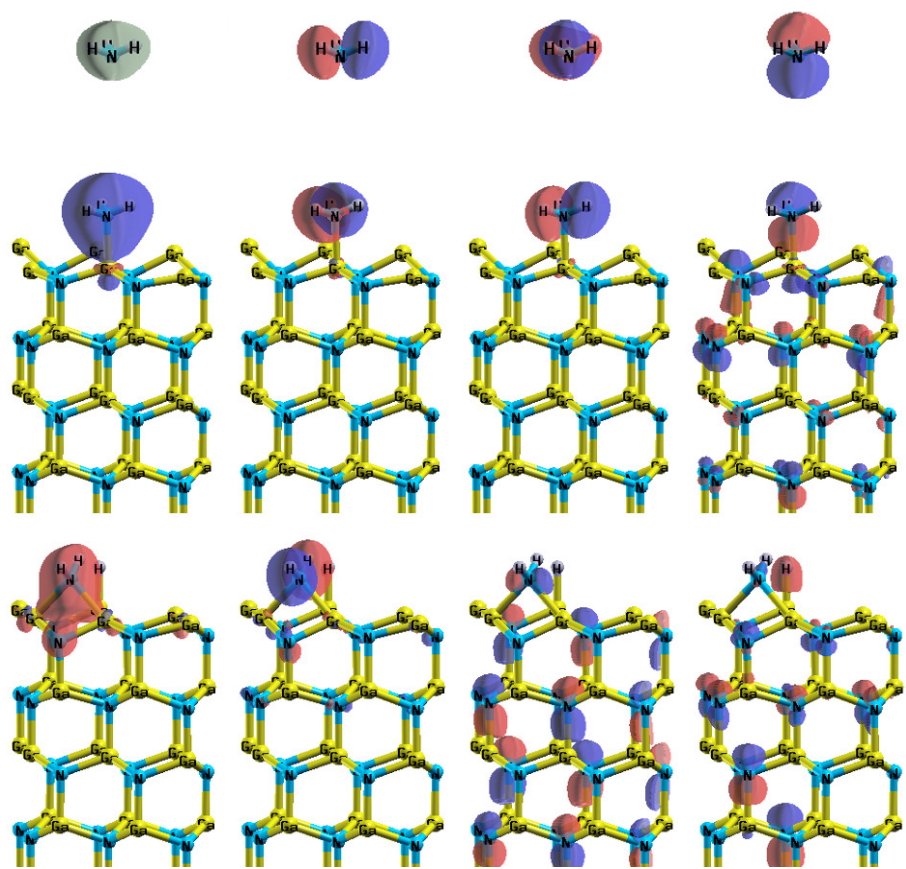

Fig. 2. Plots of the highest occupied states of the ammonia molecule: top - at far distance; middle - attached in molecular form; bottom - dissociated at the clean GaN(0001)

surface (the blue/red colors denote different sign of the function). The plots present the states with the increasing energy, from the left to the right consecutively.[35]

The basic features of the ammonia adsorption at clean GaN(0001) surface may be deduced from Fig. 3. As it is shown, at clean GaN(0001) surface, the Fermi level is pinned by Ga broken bond state located about 0.5 eV below conduction band minimum (CBM). The state has considerable dispersion, about 1.5 eV, and is therefore partially occupied. The presented here selection of the condition at the termination surface of the slab corresponds to the flat bands, i.e. no excess electric charge located at both surfaces.[28] For the case of 4 x 4 slab, the single or the three states after adsorption constitute small fraction of all surface sites therefore, after adsorption, the Fermi level is still pinned by the same state. Thus the geometric representation of the process is sufficiently large to assume that the change of the surface state as in the real adsorption process entails negligible fraction of the surface sites and does not affect the general electric state of the infinite surface.

The energetic features of the bonding of the ammonia far and at the clean GaN(0001) surface may be analyzed using the overlap population in the COHP plots, presented in Fig. 3. As it is shown, in the separate ammonia molecule, i.e. located at far distance from the surface, the eight states (i.e. four with the two spin values) are fully occupied, all degenerate with the valence band (VB) of GaN. The lowest energy state ($\sigma$ - s state) in the lower valence subband and the two other, identical energy states ($\sigma$ - p states) at the bottom of the upper valence subband. The highest energy state, ($\pi$ - p state) has its energy close to the valence band maximum (VBM). Thus, as shown in Fig. 2, all 8 quantum states are fully occupied in the ammonia molecule located far from the surface with their energy degenerate with the GaN valence band.

When the ammonia molecule is attached as a whole at the surface only the single, highest energy quantum state of the molecule is transformed. The other three states remain molecular-like without any dispersion with their energy shifted with respect to the band states only. They have negligible overlap with the uppermost Ga states, thus they remain unchanged after adsorption. The fourth one attains huge overlap with the nearest Ga state and therefore it is completely modified, i.e. with huge dispersion extending over the entire upper subband of valence band, i.e. it is converted into surface state. More precisely, the overlap between Ga surface topmost atom wavefunction and the nitrogen creates bonding and antibonding states.

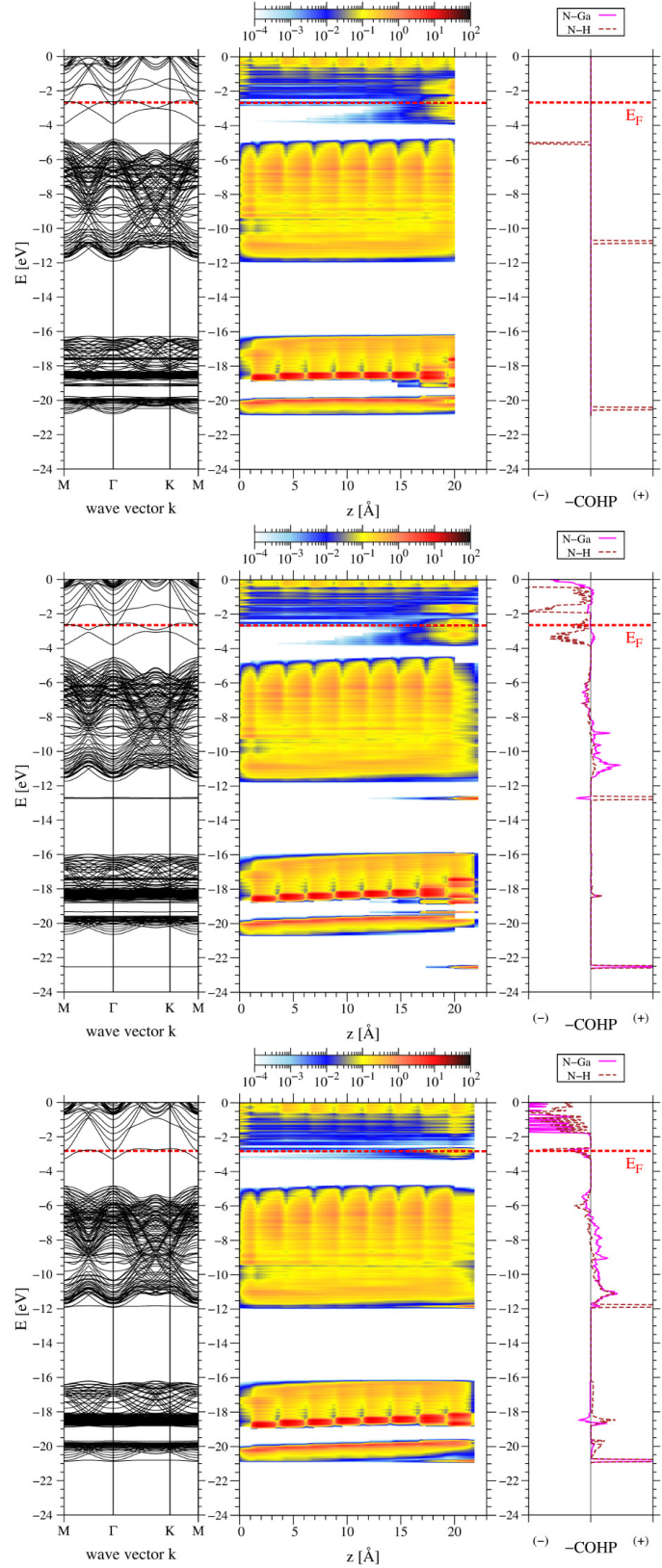

Fig. 3. Dispersion relations (left), space alignment of the bands derived from atom projected density of states (P-DOS) (middle) and Crystal Orbital Hamilton Population (COHP - right) of the N-H and Ga-N bonding of 2 x 2 GaN(0001) slab and the ammonia molecule: a) at far distance; b) attached in molecular form; c) attached dissociated. The COHP data were obtained for nitrogen-hydrogen bonds within ammonia molecule and bonds between ammonia's nitrogen located on top of the gallium atom. The positive and negative COHP values denote bonding and antibonding overlaps of the atomic states (the red and magenta color denote overlap of the states of the ammonia's nitrogen atom with the hydrogen and gallium atoms respectively).

The high energy state resulting from antibonding overlap, between the surface and adsorbing species states is located deeply in the conduction band and not identified here, the bonding state is present as dispersive state, degenerate with valence band. These four states, the three molecular and the one dispersive, remain doubly occupied, thus no charge transfer to or from these states is possible. Before and after adsorption, the Fermi level is pinned by the surface state, due to gallium non-saturated orbitals of their energy at about 0.5 eV below conduction band minimum (CBM). The charge transfer during adsorption is small as can be deduced from small slope of the band diagram.

In the case of dissociative adsorption the two molecular are transformed into dispersive surface states loosing its molecular form. In addition the two molecular states remain intact, occupied by the two electrons out of the total 8 electrons of the ammonia molecule. The dissociated hydrogen adatom is attached to the surface attaining the electronic charge from the saturated Ga states. Thus the charge transfer to Ga adatom is possible.

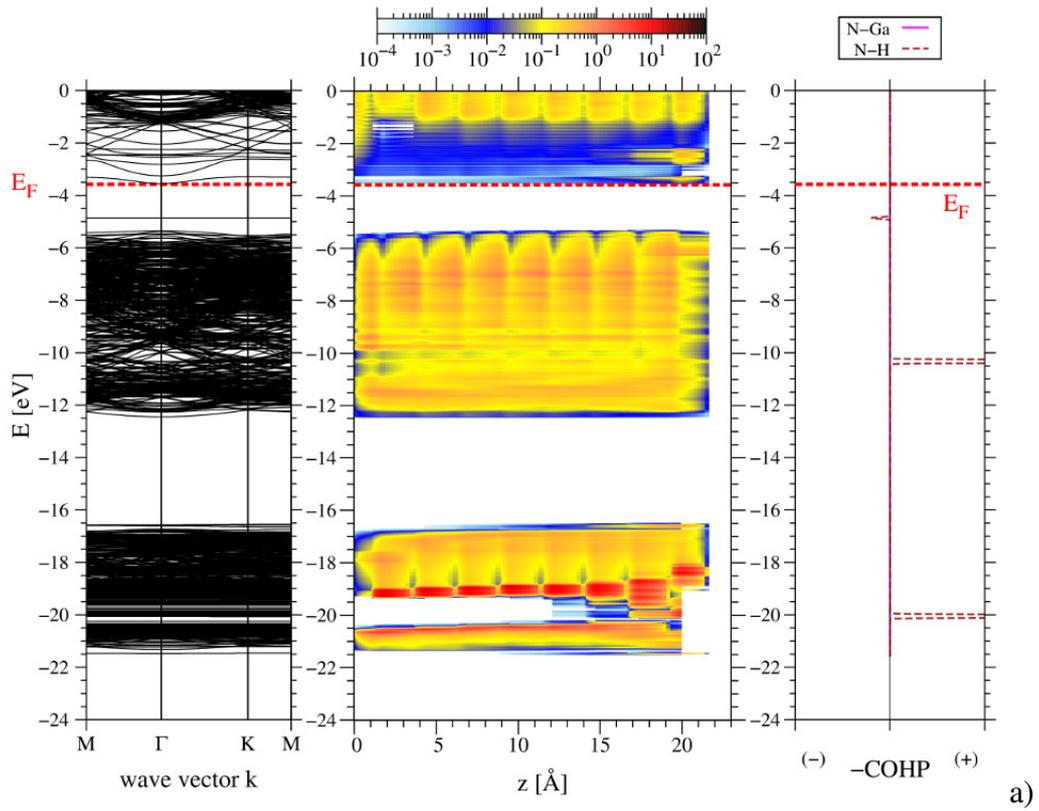

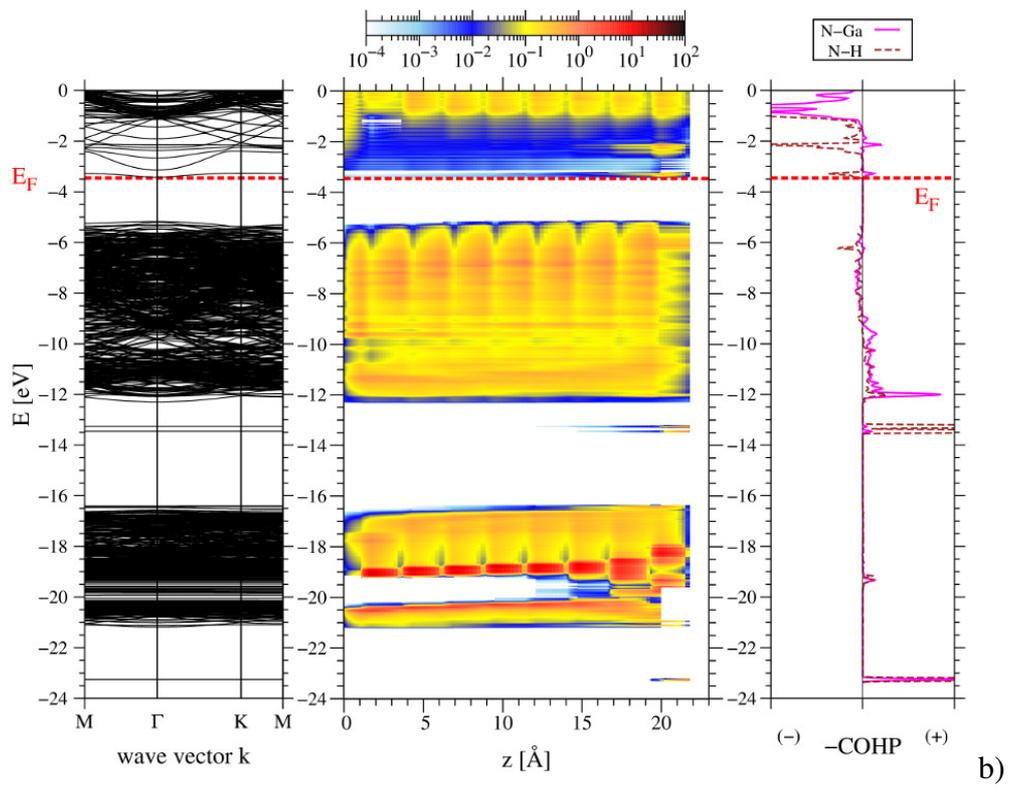

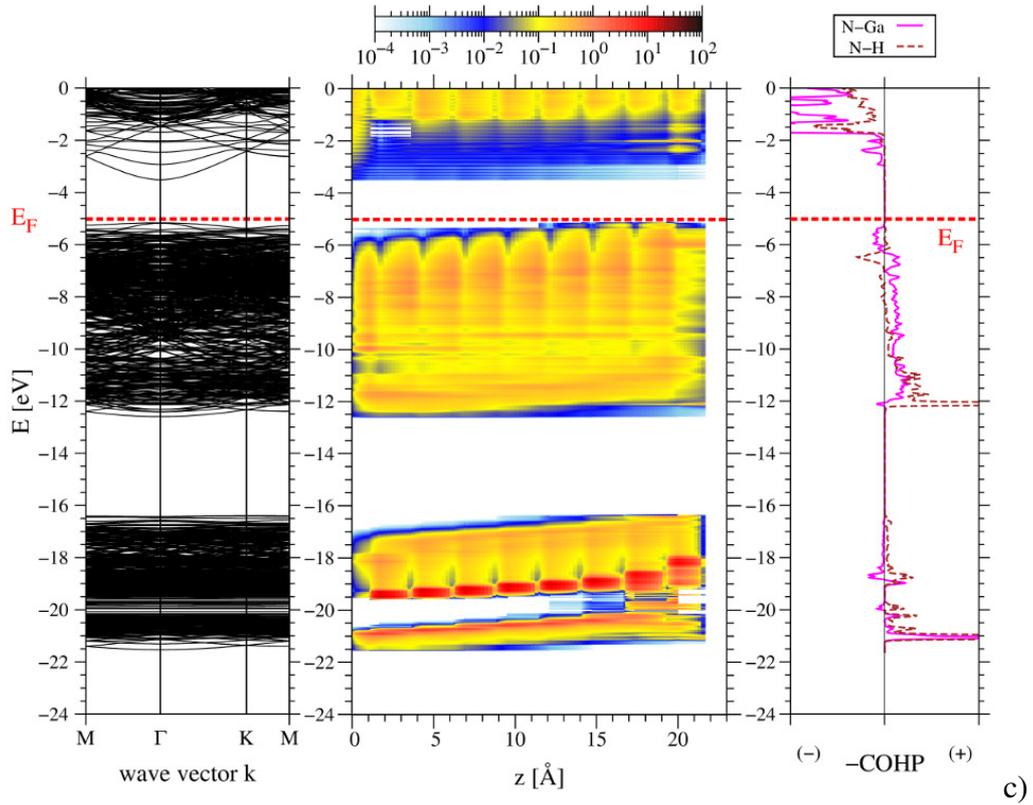

Fig. 4. Dispersion relations (left), space alignment of the bands derived from atom projected density of states (P-DOS) (middle), and Crystal Orbital Hamilton Population (COHP - right) of the 4 x 4 GaN(0001) slab with the 11 H adatoms (0.6875 ML coverage) and the ammonia molecule: a) at far distance; b) attached in molecular form; c) attached dissociated.

The 4 x 4 slabs allow investigation of the partial coverage of the surface in 1/16 ML = 0.0625 ML intervals, starting from the low coverage where the dissociation of the ammonia molecule and creation of bridge configuration dominates to the high coverage where the molecular adsorption is prevailing. The second considered case corresponds to 0.6875 ML coverage, i.e. 11 hydrogen atoms attached in the on-top positions above topmost Ga atoms, presented in Fig. 4. Before and after molecular adsorption, the Fermi level is pinned by Ga broken bond state. On the contrary, the dissociative adsorption into $NH_2$ radical and H adatoms saturates three Ga broken bonds, leaving only two uncovered. As expected, the Fermi level is shifted down to valence band maximum (VBM), standard for highly H-covered surface. Since the H-Ga state is degenerate with valence band, the Fermi level is pinned to VBM.[14]

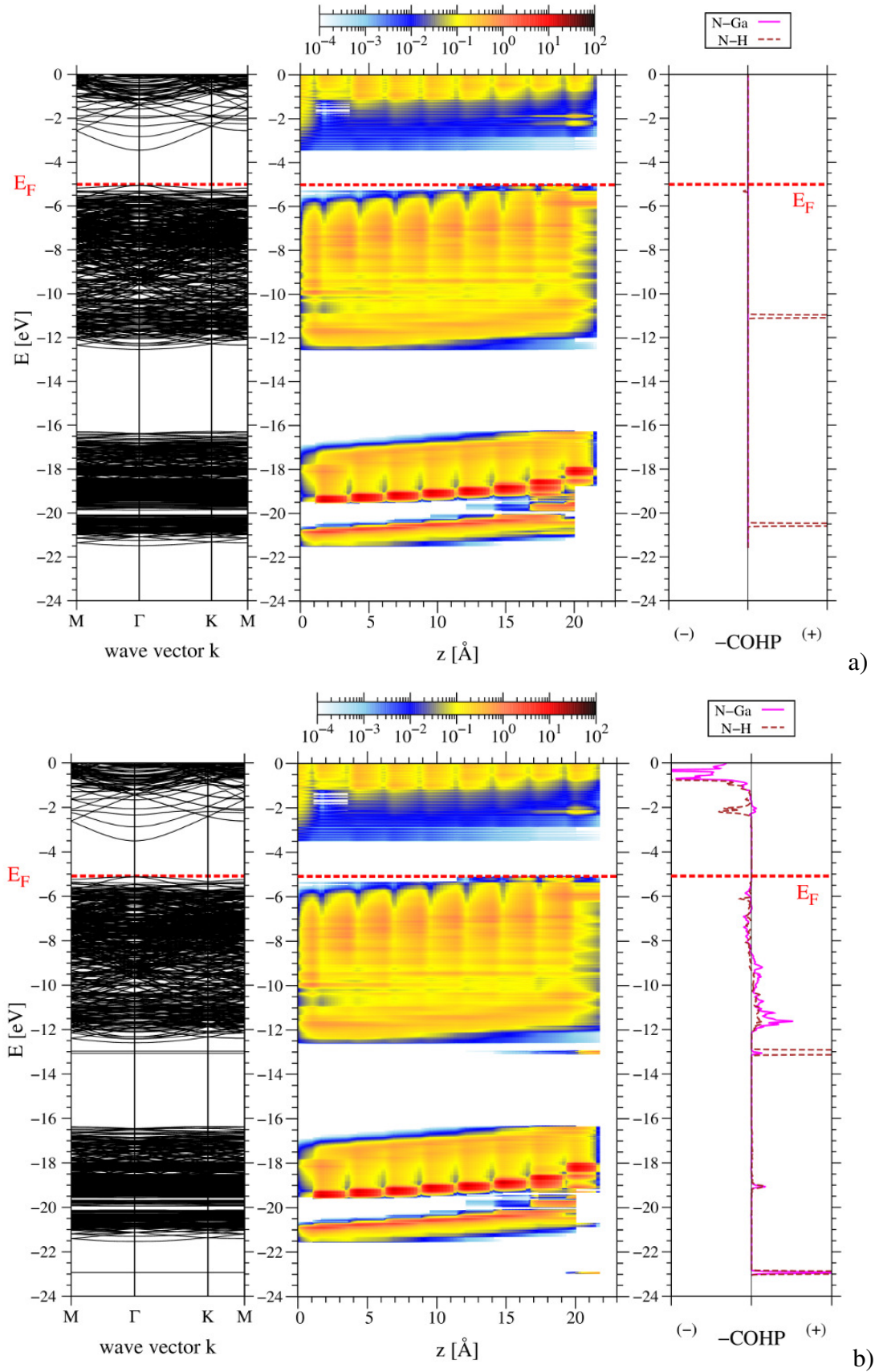

Fig. 5. Dispersion relations (left), space alignment of the bands derived from atom projected density of states (P-DOS) (middle) and Crystal Orbital Hamilton Population (COHP - right)

of the 4 x 4 GaN(0001) slab with the 14 H adatoms (0.6875 ML coverage) and the ammonia molecule: a) at far distance; b) attached in molecular form.

The COHP data indicate that the states created during adsorption are identical to those for clean surface, shown in Fig 2. Thus, Fermi level shift observed for dissociative adsorption does not affect the energy of surface states. Note however that the electric state of the surface is changed for the dissociative case. The surface is negatively charged indicating on the emergence of the acceptor state at the surface.

It is instructive to consider very high hydrogen coverage, when the number of free sites is limited. The considered case corresponds to 0.875 coverage i.e. 14 hydrogen adatoms in 4 x 4 GaN slab thus only molecular adsorption was possible. The electric state of the surface, before and after adsorption is typical for highly H-covered GaN(0001) surface with the Fermi level pinned at VBM.[14]

These obtained adsorption data thus cover wide range of the coverage, corresponding to the Fermi level pinned close to CBM, not pinned to pinned again at VBM. As it was shown recently, the adsorption energy is independent on the Fermi energy (i.e. doping in the bulk) for the Fermi level pinned at the surface while this energy is highly sensitive on the doping for Fermi level nonpinned at the surface.[28] Therefore in Fig. 6 the data are presented for n- and p-type separately..

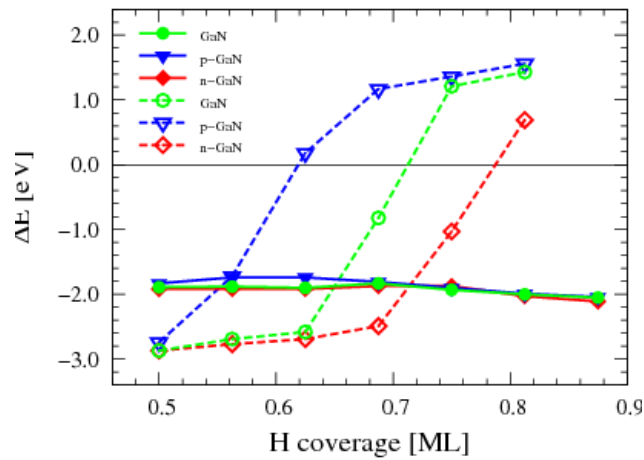

Fig. 6. Adsorption energies of ammonia molecule at relatively highly hydrogen covered GaN(0001) surface. Full and empty symbols denote molecular and dissociative process, respectively.

As shown in Fig. 5, the molecular process is characterized by the adsorption energy close to 2 eV, virtually independent of the coverage. For the molecular process the difference between n- and p-type doping is fairly small, below 0.1 eV. Note that the electric state in all presented cases did not change in the molecular adsorption processes.

In contrast to that, the dissociative process is characterized by the radical change of the energy which begins at the 0.5 ML initial coverage. It is also worth noting that the energy difference between both doping types could reach 4 eV, with the notable example of 0.6875 ML coverage, analyzed in Fig. 6, where the dissociative process is egzo- and endothermal for n- and p-type respectively. Accordingly for higher coverage, the adsorption energy is drastically reduced, and for 0.8 ML the dissociative adsorption becomes energetically unfavorable for both types. The mechanism leading to such drastic difference is related to different charge transfer in the dissociative processes as demonstrated by spatial band diagrams in Figs 3 - 5.

It is worth to stress that the difference in the adsorption energies is not related to the interaction with the neighboring atoms. The comparison of the COHP diagrams of both molecular in Fig 3b and 4b and also the dissociated configurations in Fig. 3c and 4c proves that the energies of the states are unaffected by different surrounding. Thus the difference of adsorption energies is not related to the interactions with the neighboring atoms, as postulated in standard theories of adsorbed species.[36,37]

By direct extension of the results it could be deduced that the adsorption energy of closed shell molecular species is independent of the Fermi level pinned or not and the doping in the bulk in the case when the molecular system does not disintegrates at the surface and has its states degenerate with valence band. In the case of the open shell systems, e.g. metal atoms, or in the case when molecular species disintegrate at the surface to such systems, the charge transfer leads to dependence of the adsorption energy on the pinning surface state or on doping in the bulk in the case of nonpinned Fermi level at the surface.

## V. Summary

Adsorption of the species on semiconductor surfaces was considered. It was shown that for the case of charge neutral process such as molecular closed shell systems remaining intact at the surface, the adsorption energy is unique, i.e. it is independent of the doping in the bulk or the Fermi level pinning at the surface. For the process with the charge transfer, to or from the adsorbed species, the adsorption energy depends on the on the energy of the state at the surface pinning Fermi level at the surface. As shown in Ref. 28 for the Fermi level nonpinned at the surface, the adsorption energy depends on the doping in the bulk.

The DFT simulations of ammonia adsorption at H-covered GaN(0001) surface confirmed these considerations. For molecular process, the adsorption energy is close to 2.0 eV, independent on the doping or hydrogen coverage. The dissociative process is strongly H-coverage dependent, for low H-coverage the adsorption energy is close to 2.8 eV, for high coverage it changes by more than 4 eV. Thus for low coverage the energetically preferred mode is dissociative, for high molecular, filling the vacant sited by $NH_3$ admolecules. Thus the DFT simulations prove that transition is related to the change of the Fermi level pinning from Ga- broken bond state to VBM, confirming the decisive role of charge transfer in the adsorption processes.

## Acknowledgements

The calculations reported in this paper were performed using the computing facilities of the Interdisciplinary Centre for Modelling (ICM) of Warsaw University. The research published in this paper was supported by funds of Poland's National Science Centre allocated by the decision no DEC-2011/01/N/ST3/04382. The use of XCRYSDEN shareware in generating of the atomic plots is gratefully acknowledged.[35] The research was supported in part by PL-grid infrastructure.